# Anomalous thermal activation of the electron glass dynamics in a-InO$_x$ and granular aluminum

**T. Grenet and J. Delahaye**
CNRS, Institut NEEL, F-38000 Grenoble, France
Université Grenoble Alpes, Institut NEEL, F-38000 Grenoble, France

E-mail: thierry.grenet@neel.cnrs.fr

**Abstract**:

In this article, we explore the temperature dependence of the electrical glassy dynamics in insulating amorphous indium oxide (a-InO$_x$) and granular Al films. We use non-isothermal gate voltage protocols, which can reveal changes in the dynamics induced by the temperature, when logarithmic relaxations devoid of characteristic times are at work. We demonstrate that, contrary to almost 20 years of opposite belief, the dynamics of amorphous indium oxide films in the liquid helium temperature range is thermally activated, i.e. it slows down under cooling and accelerates upon heating. Amorphous indium oxide thus adds to the list of glassy disordered systems in which we already demonstrated thermal activation, which includes granular Al and amorphous Nb$_x$Si$_{1-x}$ films. Moreover, measurements up to 40 K in a-InO$_x$ and granular Al films reveal the close similarity between the two systems and a very anomalous character of the thermal activation, with an effective activation energy increasing with $T$ as $T^2$. We so far have no explanation for it. Its further study and understanding may be important for the physics of electron glasses.

## 1. Introduction

The electron glass concept was theoretically forged in the early 80' for systems where electrons acquire a glassy dynamics characterized by macroscopically long relaxation times and thus an inability to reach equilibrium in experimental times [1]. A basic ingredient for the glassy state is a strong quenched atomic disorder, which localizes the electrons and correspondingly degrades the screening of coulomb repulsions.

On the experimental side, specific non equilibrium phenomena, interpreted in the framework of the electron glass, have been found in a number of insulating disordered systems: discontinuous Au [2, 3], amorphous and microcrystalline indium oxide [4, 5], granular Al [6, 7] and Be [8], amorphous Nb$_x$Si$_{1-x}$ [9, 10], Ge-Te alloys [11] and thallium oxide [12]. Typical of these systems (except discontinuous Au) is a logarithmic decrease of the electrical conductance with time after a quench from high to cryogenic temperature. In MOSFET-like structures where the electron glass is the channel, a so-called memory dip (MD) can be imprinted in the conductance versus gate voltage curve by keeping the system for some time at a given $V_g$ value (see for instance MDs in Figure 2 below). It reveals the partial equilibration achieved logarithmically with time, under the visited $V_g$ value, and is progressively erased once $V_g$ is changed.

Various gate voltage protocols have been used to study the features of the glassy dynamics. Recently the end of the logarithmic relaxation regime was observed in indium oxide films having low

# Anomalous thermal activation of the electron glass dynamics in a-InO$_x$ and granular Al

enough charge carrier densities and resistance values approaching the metallic state [13], which strengthens the "intrinsic electron glass" interpretation of the slow relaxations [14] against the "extrinsic" one [2, 7, 15]. Nevertheless, all studies of the slow dynamics in other systems and most of the large body of studies in indium oxide have been performed in the logarithmic regime of relaxations. The absence of characteristic time in this regime as well as the lack of an established quantitative understanding of the conductance changes, which strongly vary with temperature, caused some confusion in the interpretation of some experiments. Even a basic question like the effect of temperature on the dynamics was not fully experimentally settled. For indium oxide and granular Al, the MD dynamics first seemed to be insensitive to temperature [7] or was even claimed to accelerate when $T$ is lowered in the case of low doped indium oxide [16]. But the protocols used to reach these conclusions were later criticized [17, 18]. Using more reliable non isothermal protocols, it was instead showed that in a-Nb$_x$Si$_{1-x}$ [9] and granular Al [19] the MD dynamics is activated by temperature: MDs partially "freeze" upon cooling i.e. their erasure is slowed down when $T$ is reduced, while it is accelerated when $T$ is increased. Moreover, recent results have shown that the glassy features do not only exist at cryogenic temperatures but extend up to room temperature [20]. This prompts us to expend the temperature range over which these non-isothermal protocols are implemented.

Indium oxide is the most studied prototype of electron glasses and it is of prime importance to fully characterize its dynamics and compare it to the other systems. In this paper we investigate the effect of temperature on the glassy dynamics in the logarithmic regime. We first show that exactly like in other systems, and contrary to all previous claims, the glassy dynamics in a-InO$_x$ "freezes" upon cooling, the time needed to erase a MD becoming much longer if the sample is cooled after its formation. Moreover, using an extended protocol we show that the dynamics in indium oxide has a non-trivial dependence on the temperature history and is not a simple Arrhenius behavior. We also discover the same striking phenomenon in granular Al films.

In section 2 we present the samples and experimental methods. The results of this study are shown in details in section 3, and discussed in section 4.

## 2) Experimental methods

### 2-1) *Samples and measurement techniques.*

Amorphous indium oxide thin films were e-beam evaporated in a controlled oxygen pressure on oxidized doped Si substrates to allow for field effect measurements. In order to investigate the dynamics in a broad temperature range, we give results involving films of markedly different insulating character. Their main parameters are given in Table 1. Sample InOx1 was introduced in the cryogenic system two days after evaporation. Samples InOx2 and InOx3, initially too resistive, were gently annealed in air before insertion in the cryogenic environment. The measurements extended over several months. Each sample was kept in the dark and in vacuum or He gas during this period. As indicated in Table 1, for sample InOx3 the measurement run was interrupted by a summer break, during which the sample warmed up to room temperature and its 4K resistance per square increased from 1.7 GΩ to 5 GΩ at 4K (a change which corresponds to a rather small increase of the insulating character). For comparison with the literature, we give rough estimates of the carrier concentrations as defined and determined in [21] using the MD width. These samples lie well in the logarithmic regime of relaxation. The results obtained with a-InO$_x$ are compared to those of granular aluminum, of which three films were measured. We give in Table 1 the main parameters of the samples, and their $R(T)$ curves in Figure 1. Note that a low temperature extrapolation of the curve for sample InOx1 gives $R$(4K) of order $10^{18}$ ohms: it is significantly more insulating than all a-InO$_x$ samples previously studied in the literature.

| Sample | Sample thickness and evaporation conditions | Sample resistance per square | SiO2 gate insulator thickness | Memory dip width (in gate voltage units) and |

# Anomalous thermal activation of the electron glass dynamics in a-InO$_x$ and granular Al

|  |  |  |  | estimates of carrier densities |
|---|---|---|---|---|
| InOx1 | 100 Å<br>1 Å/s<br>P$_{O2}$ = 2.10$^{-4}$ mbar | 510 kohm @ 300 K<br>100 Gohm @ 15 K | 500 nm | 1.2 V @ 17K<br>n ≈ 10$^{21}$ cm$^{-3}$ |
| InOx2 | 100 Å<br>1 Å/s<br>P$_{O2}$ = 2.3-2.4.10$^{-5}$ mbar | 80 kohm @ 300 K<br>250 Mohms @ 4.2 K | 100 nm | 0.5 V @ 4.2K<br>n ≈ 2.10$^{20}$ cm$^{-3}$ |
| InOx3 | 100 Å<br>1 Å/s<br>P$_{O2}$ = 3.10$^{-5}$ mbar | 57 kohm @ 300 K<br>1.7 Gohms @ 4.2 K<br>then<br>5 Gohms @ 4.2 K | 100 nm | 0.25 V @ 4.2K<br>n ≈ 10$^{20}$ cm$^{-3}$ |
| GranAl1 | 100 Å<br>2 Å/s<br>P$_{O2}$ = 1.6-1.7 10$^{-5}$ mbar | 45.2 kohms @ 300 K<br>8.52 Mohms @ 4.2K | 100 nm | 1.5 V @ 4.2K |
| GranAl2 | 200 Å<br>2 Å/s<br>P$_{O2}$ = 2.10$^{-5}$ mbar | 94.6 kohms @ 300 K<br>380 MOhms @ 4.2 K | 100 nm | 1.5 V @ 4.2K |
| GranAl3 | 100 Å<br>2 Å/s<br>P$_{O2}$ = 2.6-2.8 10$^{-5}$ mbar | 920 kohms @ 300 K<br>470 Gohms @ 10 K | 100 nm | 3V @ 14K |

*Table 1: main parameters of the a-InO$_x$ and granular Al films measured.*

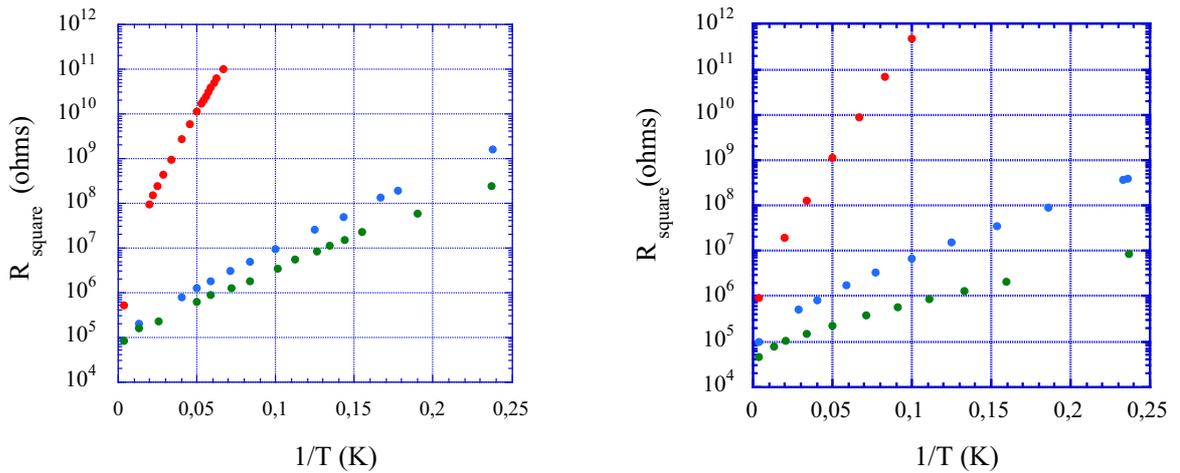

*Figure 1: resistance versus temperature curves for the a-InO$_x$ (left) and granular Al (right) samples studied.*

Measurements were performed in an evacuated insert plunged in a 100L liquid He-4 vessel, comprising a temperature-controlled sample holder. Electrical conductance measurements were

# Anomalous thermal activation of the electron glass dynamics in a-InO$_x$ and granular Al

performed in the ohmic regime using a DC bias voltage source and a Femto DLCPA-200 current amplifier. The study of the glassy dynamics involves gate voltage cycling. It was applied across the substrates' oxide layers with a Yokogawa DC voltage source.

### 2-2) Basic temperature-gate voltage protocol.

Before describing the basic protocol involved in this study, we recall that after the establishment of fixed conditions (notably constant temperature $T$ and gate voltage $V_g$), a slow evolution towards equilibrium is accompanied by a logarithmic with time decrease of the conductance. A subsequent fast $G(V_g)$ scan reveals the corresponding MD centered on the fixed $V_g$, which is a trace of the partial equilibration achieved. Setting $V_g$ to another value progressively erases the first MD and forms a new one at the new $V_g$ value.

In the regime of scale-free log($t$) relaxations, and without any precise understanding of the effect of $T$ on the relaxation *amplitudes*, isothermal measurements cannot be used to reveal any temperature dependance of MD *dynamics* [18] (see also [17]). A very natural and simple way to circumvent the difficulty is to use a two-temperatures protocol: build a MD at $T_{build}$ under $V_{gbuild}$ for a given time $t_{build}$, then set the temperature to $T_{eras}$ and the gate voltage to $V_{geras}$, and finally monitor the erasure of the $V_{gbuild}$ memory dip, using fast $G(V_g)$ scans repeated at regular time intervals (see for example Figure 2). If the MD dynamics is activated, one simply expects the erasure to be faster if $T$ is increased ($T_{eras} > T_{build}$), and slower if it is decreased ($T_{eras} < T_{build}$). Such a protocol was used in a-Nb$_x$Si$_{1-x}$ and granular Al [9, 19]. In both cases it revealed the thermal activation of the MD glassy dynamics, which could not be seen in isothermal experiments. To our knowledge this two temperatures protocol was never applied to the a-InO$_x$ electron glass. We describe here its implementation and results with this system.

Note that the cooldowns from $T_{build}$ to $T_{erase}$ were always fast compared to the building and erasure steps, typically involving one or two minutes compared to several hours. After each cycle, the sample was heated to a high enough temperature under a distant $V_g$ to erase completely its memory. The $G(V_g)$ curves consist of a stable background and MDs superposed to it. In highly resistive a-InO$_x$ samples, the stable background variation with $V_g$ is much larger than the MD amplitude, hence for each cycle the background was first measured once $T_{build}$ was established and before $V_g$ was switched from the distant value to $V_{gbuild}$. This allowed background removal when necessary.

## 3) Experimental results:

### 3-1) Evidence for thermal activation in the a-InOx electron glass.

In Figure 2 we show the erasure dynamics at $T_{erase} = 4.2$K of a MD in sample InOx3 when built at 4.2K, 6K and 8K. Comparing the sets of $G(V_g)$ and the corresponding erasure curves (fading MD amplitude versus reduced time) it is readily seen that when formed at a $T_{build}$ higher than $T_{erase}$, the MD is much more robust to erasure. One can define an erasure time extrapolating the erasure curves to zero. In the isothermal case it is well known and shown in Figure 2 that a $ln\left(1 + \frac{t_{build}}{t}\right)$ dependence is obeyed [7, 22], giving an erasure time equal to $t_{build}$. Would the MD dynamics be insensitive to $T$, the same erasure time would be obtained in the non-isothermal cases. Instead, one gets erasure times larger by orders of magnitude when $T_{build} > T_{erase}$, as can be seen in Figure 2.

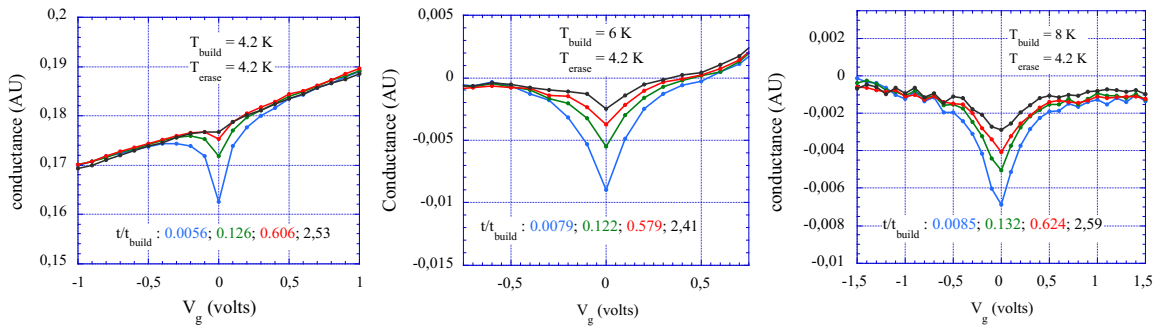

# Anomalous thermal activation of the electron glass dynamics in a-InO$_x$ and granular Al

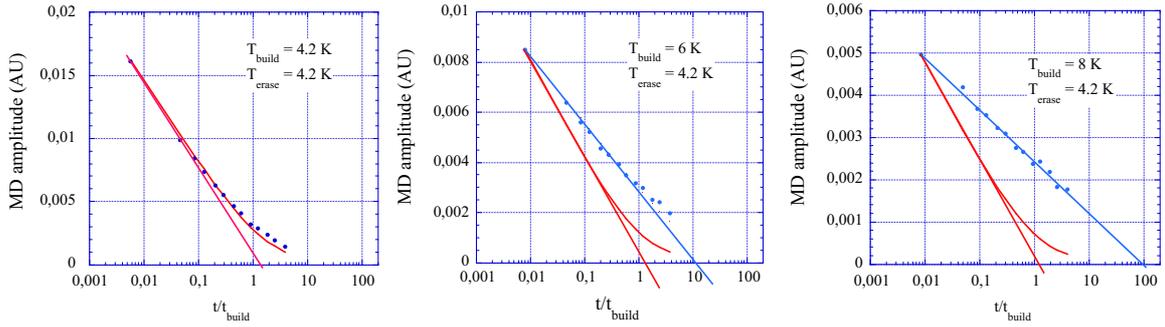

*Figure 2: erasures at T$_{erase}$ = 4.2K of memory dips built at three different T$_{build}$ in sample InOx3. Top panel: selections of G(V$_g$) scans obtained at similar $\frac{t}{t_{build}}$ values, showing the progressive erasure of the previously formed MD. Here V$_{gerase}$ = -2V. Building times are respectively 5 hours, 3.5 hours and 3.25 hours. Left: raw curves are shown, middle and right: the background was subtracted as explained in the main text. The shape of the background is discussed in a next section. Bottom panel: time dependence of the fading MD amplitudes (blue points and extrapolation), compared to the expected isothermal behaviour (red curves). The MD amplitude is taken as $mean\{G(-0.5V), G(+0.5V)\} - G(0V)$.*

As explained in [19] the acceleration of the MD erasure when T$_{erase}$>T$_{build}$ is more difficult to reveal. But it can be visualized by applying a short higher temperature excursion during the erasure step of an otherwise isothermal experiment. The result is a clear acceleration of the MD erasure, as illustrated in Figure 3.

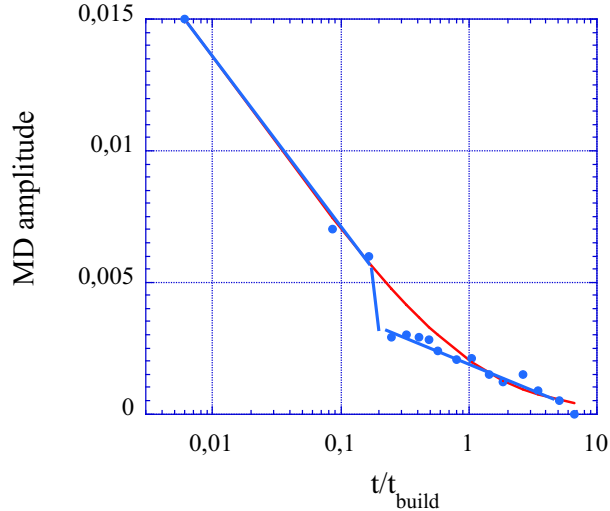

*Figure 3: erasure curve obtained with sample InOx1 during an isothermal protocol at T$_{erase}$=T$_{build}$ =36K, except for a short excursion at 40K during the erasure step. The excursion causes a clear acceleration of the MD erasure compared to the expected purely isothermal curve. In this example: t$_{build}$ = 13,75 hours at T$_{build}$ = 36K. Erasure performed at T$_{erase}$ = 36K except for the excursion at 40K started after the third measured point: 10 mins to rise and stabilize from 36K to 40K, 20 mins at 40K, and 5 mins to get back to 36K. The MD amplitude is reduced by a factor 2 by the temperature excursion. This cannot be due to a thermal broadening of the MD, the relative temperature change being modest. The accelerated erasure during the excursion at 40K results in a local depletion of the relaxation time distribution once back to 36K, hence the slowdown of the erasure until it merges back with the expected purely isothermal curve.*

These results show for the first time that cooling freezes the MD dynamics in the a-InO$_x$ electron glass. This is the first main result of the paper. This finding motivates a broader study of the dynamics, involving higher temperatures and changing both T$_{build}$ and T$_{erase}$.

# Anomalous thermal activation of the electron glass dynamics in a-InO$_x$ and granular Al

### 3-2) *Evidence for highly non-Arrhenius behaviour in the a-InOx and granular Al electron glasses*.

Assuming a simple Arrhenius behaviour of the MD slow dynamics, one may expect that in a non-isothermal experiment with $T_{build} \neq T_{erase}$, the erasure time is given by: $t_{erase} = t_{build} \times exp\left(\frac{E_a}{k_B} \times \frac{(T_{build}-T_{erase})}{T_{build}T_{erase}}\right)$ where $E_a$ is an activation energy [19]. Analyzed along this line the data of Figure 2 imply $E_a \approx 32 \pm 4K$, close to the value obtained in a similar manner in granular Al [19].

However, we found that experiments with different $T_{erase}$ values uncover a more complicated situation. As an example, we show in Figure 4 the case where $T_{build}$ = 16K and $T_{erase}$ = 14K. First note that the isothermal erasure curve at $T_{build} = T_{erase}$ =14 K is the same as at 4.2 K. This was checked at all the temperatures used and illustrates that as long as the glassy relaxation is in the logarithmic regime with experimentally unsurpassable relaxation times, the $T$ dependence of the dynamics is not revealed by isothermal protocols. More importantly, note that if the activation energy was of order 32 K, within the scatter of the points the two curves in Figure 4 would be hardly discernable. Instead, they are well separated and one gets an activation energy more than ten times larger than previously: $E_a \approx 370K \pm 15\%$.

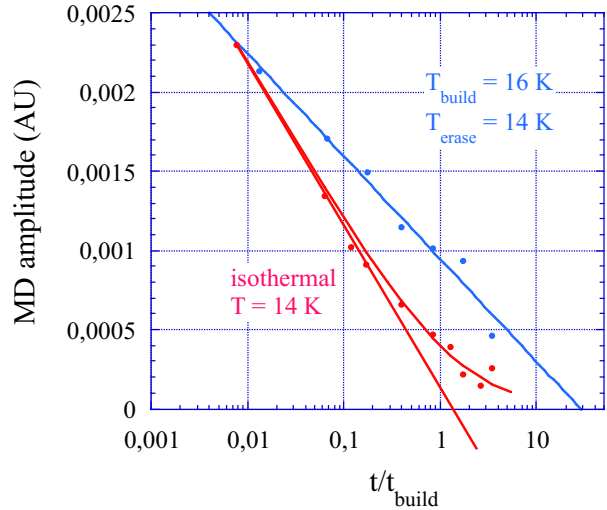

*Figure 4: comparison of erasure curves obtained with sample InOx3 for an isothermal protocol ($T_{build} = T_{erase}$ =14 K) and a non-isothermal one ($T_{build}$ = 16 K, $T_{erase}$ =14 K). Assuming a simple Arrhenius behaviour, an activation energy ~370 K is extracted, i.e. ten times larger than the value estimated from the measurements of Figure 2 (see main text for details).*

This shows that the MD dynamics is not simply activated but has a more complex dependence on the thermal history. The activation energies extracted from the data are effective values which we use to quantify departures from simple Arrhenius.

For a systematic investigation we vary $T_{build}$ and $T_{erase}$ and extract as above the effective activation energy $E_a(T_{build},T_{erase})$. We show below the result obtained with the three indium oxide samples of Table 1. The measurements involve temperatures up to 40K and erasure times up to $10^5 t_{build}$.

We first vary $T_{build}$ and $T_{erase}$ keeping their difference relatively small. Thus the measurements, although not isothermal in order to be able to reveal the freezing and $E_a$, may be considered performed at a relatively well defined temperature close to $\frac{T_{build}+T_{erase}}{2}$. We plot below $E_a(\frac{T_{build}+T_{erase}}{2})$ and observe a strong temperature dependance of, ruling out by far a simple Arrhenius law. Instead, a behaviour close to $E_a \propto \left(\frac{T_{build}+T_{erase}}{2}\right)^{1.8}$ (straight line) is observed, with $E_a$ rising over 2000 K for temperatures approaching 40K.

**Anomalous thermal activation of the electron glass dynamics in a-InO$_x$ and granular Al**

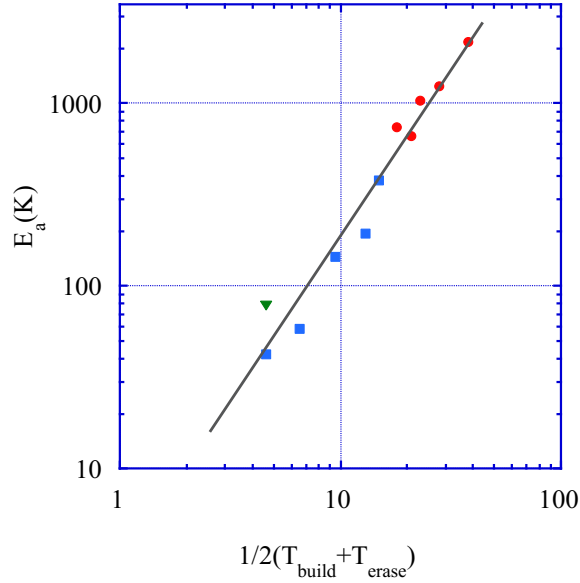

*Figure 5: dependence on the average temperature $\frac{T_{build}+T_{erase}}{2}$ of the effective activation energy $E_a$ of the MD dynamics in a-InO$_x$ films, extracted from non-isothermal experiments (see text for details). The strong temperature dependence corresponds to a slope close to 1,8. Involved ($T_{build}$,$T_{erase}$) couples in Kelvin are: (5,4.2), (7,6), (10,9), (14,12), (16,14), (19,17), (22,20), (24,22), (30,26) and (40,36). Samples: red circles = InOx1, green triangles = InOx2, blue squares = InOx3.*

The same experiments were conducted with granular Aluminum films which details are given in Table 1 and Figure 1. A very similar behaviour was observed, the measured effective activation energies are shown in Figure 6 using the same scales as in Figure 5 for ease of comparison.

This surprising non-Arrhenius behaviour is the second main finding of our study.

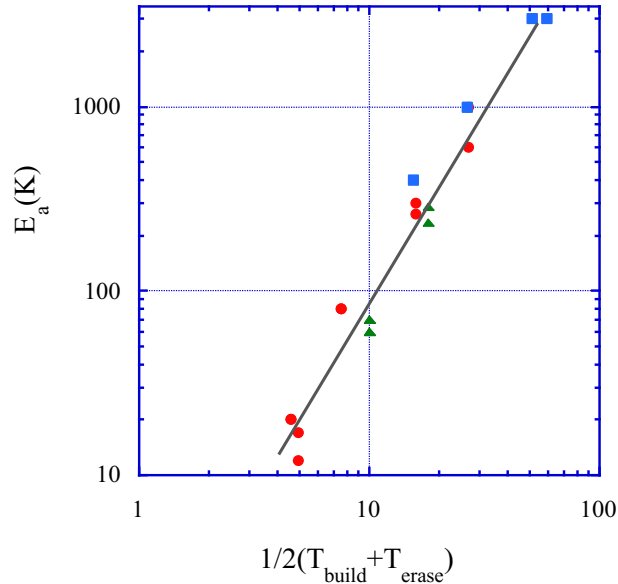

*Figure 6: dependence on the average temperature $\frac{T_{build}+T_{erase}}{2}$ of the effective activation energy $E_a$ of the MD dynamics in granular Al films, extracted from non-isothermal experiments (see text for details). The strong temperature dependence corresponds to a slope close to 2. Involved ($T_{build}$,$T_{erase}$) couples in Kelvin are: (4.9,4.2), (5.6,4.2), (8,7), (11,9), (16,15), (17,15), (19,3,16,7), (27,26), (28,26), (52,50) and (60,58). Samples: red circles = granAl1, green triangle = granAl2, blue squares = granAl3.*

# Anomalous thermal activation of the electron glass dynamics in a-InOx and granular Al

### 3-3) Thermal activation and frozen $G(V_g)$ backgrounds.

The two temperatures protocol is the simplest and clearest one we can think of to demonstrate thermal activation. But at temperatures higher than 40-60K it becomes difficult to implement as the MDs become wide compared to the accessible $V_g$ range and small in relative amplitude (unless the resistance of the films is significantly increased). However thermal activation can also be discerned in the low temperature $G(V_g)$ backgrounds.

We recently found that MDs can be observed up to room temperature in all the systems we investigated, including a-InOx films [20], which suggests that the trend observed in Figures 5 and 6 continues up to room $T$. Hence when a sample is cooled from room $T$ to cryogenic temperatures under a constant $V_{g\,cooling}$, we expect that MDs of progressively narrower thermal width are continuously frozen, which results in a quenched background centered on $V_{g\,cooling}$. Actually, it is the existence of this broad quenched background which first prompted us to search for thermal activation of the MD dynamics in NbSi [9]. We show in Figure 7 that such a feature indeed also exists in indium oxide. After a cooldown to 4,2K under $V_{g\,cooling} = 0$ V, a broad V-shape background is present centered on this value, and is essentially unsensitive to subsequent gate voltage changes at low $T$. As expected, zero gate voltage has nothing special. A cool-down under another $V_{gcooling}$ value produces a shifted V-shape background. Note that the frozen background is formed during the cooling and not at a well-defined temperature. Its low $T$ relaxation under a $V_g$ change cannot be easily compared with that of a $T_{build}$ MD.

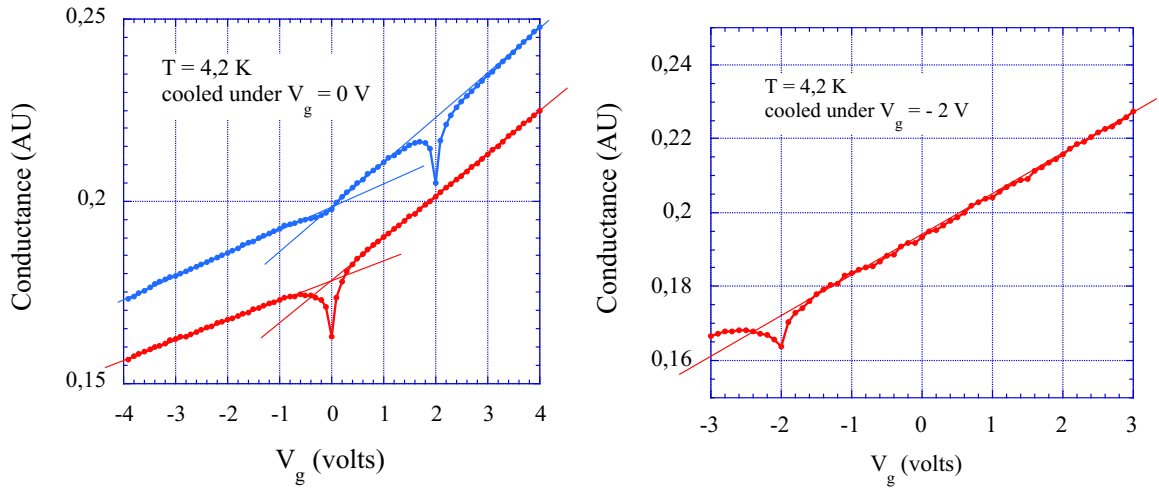

*Figure 7: Left: broad Vg scans measured after sample InOx3 was cooled from T≈ 90 K to 4,2 K under Vg = 0 V and kept under that value for 5 hours (lower curve), and after Vg was subsequently set to +2V for more than a day (upper curve). The narrow low temperature contribution to the MD is displaced to +2V while the broader background remains unchanged. Right: broad Vg scan measured after sample InOx3 was cooled to 4,2 K under Vg = -2 V. The broad V-shape background is now centered on -2V. It is thus not an intrinsic equilibrium feature, and is formed during the cooling, as expected for the continuous freezing of MDs.*

### 4) Discussion:

The data presented above first show unambiguously that the temperature influences to a great extent the memory dip dynamics in the indium oxide electron glass. Even a modest relative change of $T$ after the growth of a new MD alters significantly the time needed to subsequently erase it, a phenomenon we had already uncovered in other electron glasses. We here establish that indium oxide is no exception.

Second, we find an intriguing non-Arrhenius behaviour, which manifests in a strong temperature dependance of the effective activation energy extracted from the measurements. When the two temperatures $T_{build}$ and $T_{erase}$ are relatively close to each other and hence the experiment can be

# Anomalous thermal activation of the electron glass dynamics in a-InO$_x$ and granular Al

characterized by an average temperature $T_{average} = \frac{1}{2}(T_{build} + T_{erase})$, one gets $E_a \propto (T_{average})^\varepsilon$ with $\varepsilon$ close to 2. This phenomenon is shared by a-InO$_x$ and granular Al, irrespective of their resistance per square. Note that for nearby $T_{build}$ and $T_{erase}$ values, $E_a \propto (T_{average})^2$ implies that $\ln\left(\frac{t_{erase}}{t_{build}}\right) \propto \Delta T = (T_{build} - T_{erase})$. Hence the observed time ratio $\frac{t_{erase}}{t_{build}}$ is solely determined by $\Delta T$ and does not depend on the temperature range. The observation of an effective activation energy decreasing when $T$ is decreased is opposite to what to could be expected if a glass transition was approached.

In a previous work [19] data pertaining to granular Al could be quite well fitted to simple Arrhenius (hence with a single activation energy). But in that case only $T_{build}$ was changed while $T_{erase}$ was kept at 4.2 K. This prompted us to get more insight on the relative roles of both temperatures, so we repeated the experiments by varying one while keeping the other constant. Results of such experiments are shown in Figure 8 and Figure 9 for both systems.

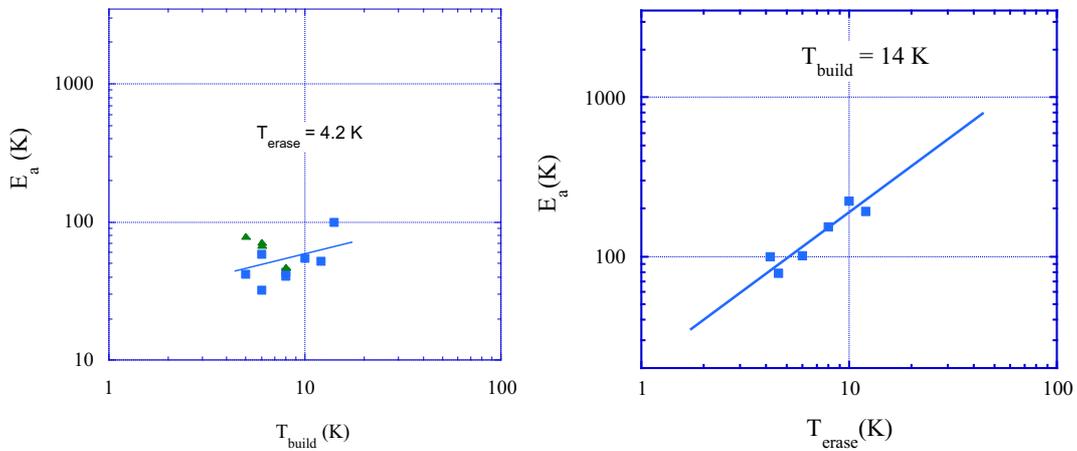

*Figure 8: dependance on $T_{build}$ (resp. $T_{erase}$) of the effective activation energy $E_a$ of the MD dynamics in a-InO$_x$ films, extracted from non-isothermal experiments of the MD dynamics, when $T_{erase}$ (resp. $T_{build}$) is kept constant. Involved ($T_{build}$,$T_{erase}$) couples in Kelvin are: left (5,4.2), (6,4.2), (8,4.2), (10,4.2), (12,4.2) and (14,4.2); right (14,4.2), (14,4.6), (14,6), (14,8), (14,10) and (14,12). Samples: green triangles = InOx2, blue squares = InOx3.*

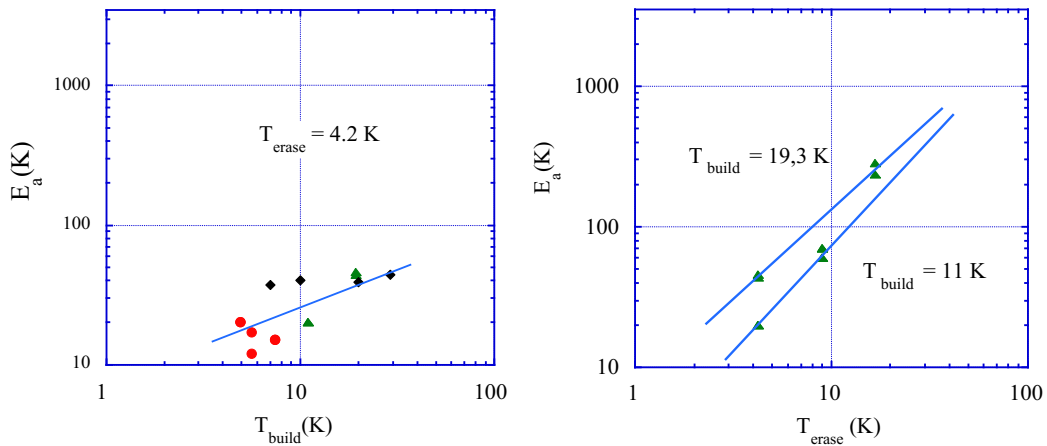

*Figure 9: dependance on $T_{build}$ (resp. $T_{erase}$) of the effective activation energy $E_a$ of the MD dynamics in the granular Al films of Table 1, extracted from non-isothermal experiments when $T_{erase}$ (resp. $T_{build}$) is kept constant. Left: $E_a$ values extracted from the data of Figure 5 in [19] are also shown. Involved ($T_{build}$,$T_{erase}$) couples in Kelvin are: left (4.9,4.2), (5.6,4.2), (7,4.2), (7.4,4.2), (10,4.2), (11,4.2), (19.3,4.2), (20,4.2) and (29,4.2); right (11,4.2), (11,9), (19.3,4.2) and (19.3,16.7). Samples: red circles = granAl1, green triangles = granAl2, black diamonds = sample from [19].*

# Anomalous thermal activation of the electron glass dynamics in a-InO$_x$ and granular Al

A clear trend is observed with the erasure temperature: $E_a$ increases like $T_{erase}$ with InO$_x$ and even faster in the case of granular Al. As for the effect of $T_{build}$ alone, a much smaller positive dependance is observed, which is hard to quantify due to the scatter of the data. The smaller dependance may explain the results of [19]. A tentative quantitative description of our whole set of data may be $E_a \propto T_{build}^{\alpha} T_{erase}^{\beta}$ with $\alpha < \beta$ and $\alpha + \beta \approx 2$. While compatible with our results within the present uncertainties, a more precise knowledge of the effect of $T_{build}$ is necessary to confirm or invalidate it.

The present findings contradict previous claims (and associated interpretations) that the MD dynamics is temperature independent in the a-InO$_x$ electron glass for highly doped samples (hence samples in the logarithmic regime) [16]. We have already pointed out [18, 19] that such claims based on isothermal experiments were not reliable. Interestingly in [23] a MD formed at 4K was measured after cooling to 1.6K. Unfortunately its erasure dynamics at this low temperature was not shown. It is probable that the observed remanence of its 4K thermal width at 1.6K was a signature of freezing. As shown above, another signature of an activated glassy dynamics is the presence of a frozen broad dip in $G(V_g)$ curves centered on the gate voltage at which the sample was cooled from high (room) temperature. Such a feature superposed to an equilibrium linear contribution appears in several instances in the literature on indium oxide, examples of which can be seen in [24]. However, when considered it was generally thought to be a "thermodynamic" equilibrium property and was discarded from discussions of the glass. The final shape and amplitude of the frozen feature are determined by the T dependance of the MD, and may also depend on the cooling conditions. Hence it is expected to vary among different batches of samples and systems, and may not always be prominent. For example, we previously found it to be quite ample in NbSi [9] and rather small in granular Al [19].

Comparing different systems, the question of the effect of microstructure on the glassy phenomenology arises [25]. We found that amorphous systems (a-Nb$_x$Si$_{1-x}$ and a-InO$_x$) and granular Al share a similar activated electron glass dynamics. The similitude of behaviour shown in the present study between a-InO$_x$ and granular Al is striking, like with most other aspects of the glassy phenomenology. As for a-Nb$_x$Si$_{1-x}$ the demonstration of activation only involved MDs built at 9K and 20K, all erased at 4,2K. Nevertheless, the non-Arrhenius character was obvious, the dependance of $E_a$ on $T_{build}$ being stronger than with the other systems [10]. Finally discontinuous gold films show a clear evidence of MD freezing upon cooling. The effect could be stronger than in the other systems, but as a different protocol was applied, a quantitative comparison is not possible [3]. As of now it remains difficult to establish a clear role of microstructural specificities.

We finally discuss the effective activation energies. It may not be very surprising that a simple Arrhenius approach cannot describe the whole set of data: with highly disordered systems one expects the actual activation energies associated to the electronic degrees of freedom to be broadly distributed. In [19] we supposed that the slow relaxations can be described using relaxation times of the form $\tau_{ij} \propto exp(\frac{E_{a_i}}{k_B T}) exp(\xi_j)$ where the activation energies $E_{a_i}$ and the tunneling exponents $\xi_i$ are independently and broadly distributed (flat distribution from zero to a maximum value). In the case where $\xi_{max} \gg \frac{E_{a\,max}}{k_B T}$ for all experimental temperatures, which generates the logarithmic regime of dynamics, the MD erasure is approximately simply activated with a temperature independent $E_a \approx \frac{E_{a\,max}}{2}$. Simulations show that non-flat $E_{a_i}$ distributions in the interval [0, $E_{a\,max}$] can generate a temperature dependance of the effective $E_a$ extracted from the experiments, but it does not seem possible to reproduce our main experimental features, notably the approximately quadratic increase of $E_a$ with $T_{average}$. This dependance is indeed very intriguing and we presently have no explanation for it.

The highest values for the effective activation energies $E_a(T)$ we obtained (a few $10^3$ K) correspond to a few $10^2$ meV. Such high values are still smaller than the carrier disorder potential estimated to be of a few eVs (a few $10^4$ K) in highly doped a-InO$_x$ [25]. In granular Al near the metal to insulator transition, the charging energy of the few nanometers sized Al islands surrounded by Al$_2$O$_3$ is generally estimated to be of several hundred Kelvins. Further in the insulating regime, thicker Al$_2$O$_3$

# Anomalous thermal activation of the electron glass dynamics in a-InO$_x$ and granular Al

barriers may produce somewhat larger values. Moreover if the slow relaxation phenomena involve multi-electron hops, as is generally admitted in the electron glass picture, several "individual" activation energies may add up to produce $E_a$. Hence the quite high effective activation energies we observed do not seem to be unreasonable. It is probable that the values of $E_a(T)$ reflect in some way the real activation energy ranges involved. However, it seems difficult to say more unless a better understanding of what is going on is acquired.

## 5) Conclusion:

In conclusion, in spite of long standing opposite claims, we found evidence for a thermal activation of the electron glass dynamics of indium oxide films: we showed the erasure of a memory dip formed at a given $T$, to be all the more so slow as it is realized at a lower temperature, and to be accelerated upon heating. This illustrates that, as long as the conductance relaxations are logarithmic in time, only non-isothermal gate voltage protocols can reveal thermal activation, a point that was overlooked in previous studies in this system. The same conclusions were already drawn up in our previous studies of amorphous NbSi and granular Al films. Hence the $T$ dependence of the glassy dynamics seems to be rather universal in disordered electron glasses.

Moreover, by varying the temperatures between 4.2 K and 50 K in a-InO$_x$, we could explore the phenomenon in more details. In particular we extracted an effective activation energy which goes like $T^2$ and reaches values as high as 2000 K at 40 K. Its decrease upon cooling is opposite to the approach of a glass transition. The measurements were repeated with granular Al films and lead to a strikingly similar result. This surprising non-Arrhenius activation law does not seem to depend on the system details and may possess some universality. We have no simple explanation for it. Its understanding may be of importance for the physics of coulomb glasses and clearly deserves further experimental and theoretical works.

An interesting perspective is offered by the recent observation of the end of the logarithmic relaxation in lightly doped and low resistance indium oxide films [13]. In these, a characteristic relaxation time can be determined using simple isothermal protocols. The evolution of this relaxation time with charge carrier density and films resistance has already been analysed. It would now be interesting to study its temperature dependence and compare it with our present results obtained in films of higher charge carrier density and resistance.

# Anomalous thermal activation of the electron glass dynamics in a-InO$_x$ and granular Al